\documentclass[fleqn,10pt]{wlscirep}
\usepackage[utf8]{inputenc}
\usepackage[T1]{fontenc}
\usepackage{lineno}
\usepackage{multirow}
\usepackage{pythonhighlight}
\usepackage{algorithm}
\usepackage{algpseudocode}
\usepackage{amssymb}
\usepackage{lipsum}
\usepackage{subfigure}
\usepackage{hyperref}
\usepackage{tasks}

\algnewcommand{\algorithmicand}{\textbf{ and }}
\algnewcommand{\algorithmicor}{\textbf{ or }}
\algnewcommand{\algorithmicto}{\textbf{ to }}
\algnewcommand{\OR}{\algorithmicor}
\algnewcommand{\AND}{\algorithmicand}
\algnewcommand{\TO}{\algorithmicto}
\algnewcommand{\VAR}{\texttt}

\title{Extracting the U.S. building types from OpenStreetMap data}

\author[1,*]{Henrique F. de Arruda}
\author[1]{Sandro M. Reia}
\author[1]{Shiyang Ruan}
\author[1]{Kuldip S. Atwal}
\author[2]{Hamdi Kavak}
\author[1]{Taylor Anderson}
\author[1]{Dieter Pfoser}
\affil[1]{Geography and Geoinformation Science, College of Science, George Mason University, 4400 University Dr., Fairfax, 22030, VA, U.S.}
\affil[2]{Center for Social Complexity, College of Science, George Mason University, 4400 University Dr., Fairfax, 22030, VA, U.S.}

\affil[*]{corresponding author: Henrique F. de Arruda (h.f.arruda@gmail.com)}

\begin{abstract}
Building type information is crucial for population estimation, traffic planning, urban planning, and emergency response applications. Although essential, such data is often not readily available. To alleviate this problem, this work creates a comprehensive dataset by providing residential/non-residential building classification covering the entire United States. We propose and utilize an unsupervised machine learning method to classify building types based on building footprints and available OpenStreetMap information. The classification result is validated using authoritative ground truth data for select counties in the U.S. The validation shows a high precision for non-residential building classification and a high recall for residential buildings. We identified various approaches to improving the quality of the classification, such as removing sheds and garages from the dataset. Furthermore, analyzing the misclassifications revealed that they are mainly due to missing and scarce metadata in OSM. A major result of this work is the resulting dataset of classifying 67,705,475 buildings. We hope that this data is of value to the scientific community, including urban and transportation planners. 
\end{abstract}
\begin{document}

\flushbottom
\maketitle

\thispagestyle{empty}

\section*{Background \& Summary}

Cities, towns, and villages are complex systems~\cite{bettencourt2021introduction} concerning organization and services. They serve as economic, cultural, and political centers and are central to social activities. Their characteristics can vary in population density, urban planning, and infrastructure~\cite{alvarez2014cost,glover1975effect}. Additionally, the differences among regions can be related to human mobility~\cite{reia2024function}.

To understand the organization of a city, researchers can analyze its infrastructure~\cite{domingues2018topological,de2016minimal,tokuda2021spatial}. Domingues, \emph{et al.}~\cite{domingues2018topological} have shown that cities on different continents can be distinguished by the structural properties of the road networks. Building footprints~\cite{boo2022high,ye2024enhancing} can also be used to understand the cities. For example, they can be combined with census information to determine the population of city subregions~\cite{boo2022high,ye2024enhancing}. Building footprints can also be used for estimating energy consumption~\cite{wang2022data}, urban planning~\cite{hamaina2012towards}, disaster assessment and response~\cite{kahraman2016disaster,putra2017humanitarian}, urban mobility~\cite{wu2020novel}, and mapping (e.g., land use maps~\cite{li2021mapping}, 3D models~\cite{park2019creating}, and digital twins~\cite{caldarelli2023role}).

In addition to the footprint geometry, building classification is important information in such applications but is often missing from official administrative data. Often, building footprints without the classification or land use data without the building footprints are available. While techniques exist to estimate building types by combining datasets~\cite{xu2023national}, there is no centralized repository for building types in the U.S. 

Instead of administrative data, a typical data source is OpenStreetMap (OSM)~\cite{OpenStreetMap}, which uses crowdsourcing to curate a global-scale geospatial dataset. Although focusing initially on road network data, OSM progressively includes Point-Of-Interest (POI) data and building footprints. In 2018, Microsoft generated a massive dataset of computer-generated building footprints. This dataset, covering the entire U.S., has subsequently been added to OSM \cite{MicrosoftBuildingFootprintData}.

Since many studies rely on OSM data~\cite{atwal2022predicting,domingues2022identification,costa2017mapping,zhou2022assessing}, researchers have examined its quality for research suitability. Data quality has multiple dimensions, considering completeness and accuracy of building footprint geometries and the associated annotations. Here, annotation refers to the contextual information that OSM users add to the geographic features using tags - key-value pairs used to describe the attributes. For example, the key \emph{building} can be associated with the value ``house'' to form a tag (\emph{building}: ``house''). Zhang \emph{et al.}~\cite{zhang2022assessing} assessed the completeness of OSM building footprint geometries by comparing them to population data.
Other studies compare official administrative data with OSM (e.g., ~\cite{hecht2013measuring,moradi2023evaluating}). For example, an analysis of OSM buildings in Germany collected in 2011 and 2012 showed low completeness regarding the building footprint geometries compared to official data~\cite{hecht2013measuring}. The data completeness for the Saxony was 15\% in 2011 and increased to 23\% in 2012. In another example, for Quebec, Canada, Moradi \emph{et al.}~\cite{moradi2023evaluating} found improvements in completeness and accuracy of the building footprint geometries and annotations over time.
In comparison with building geometries, annotations, or tags, in OSM data can suffer from incomplete or inaccurate information, especially in rural areas \cite{vargas2019correcting}. Furthermore, high levels of annotation completeness do not necessarily imply high accuracy, as errors or incorrect annotations can still result in poor overall data quality~\cite{barron2014comprehensive, mcgough2024more}.

\begin{figure*}[h]
    \includegraphics[width=1\textwidth]{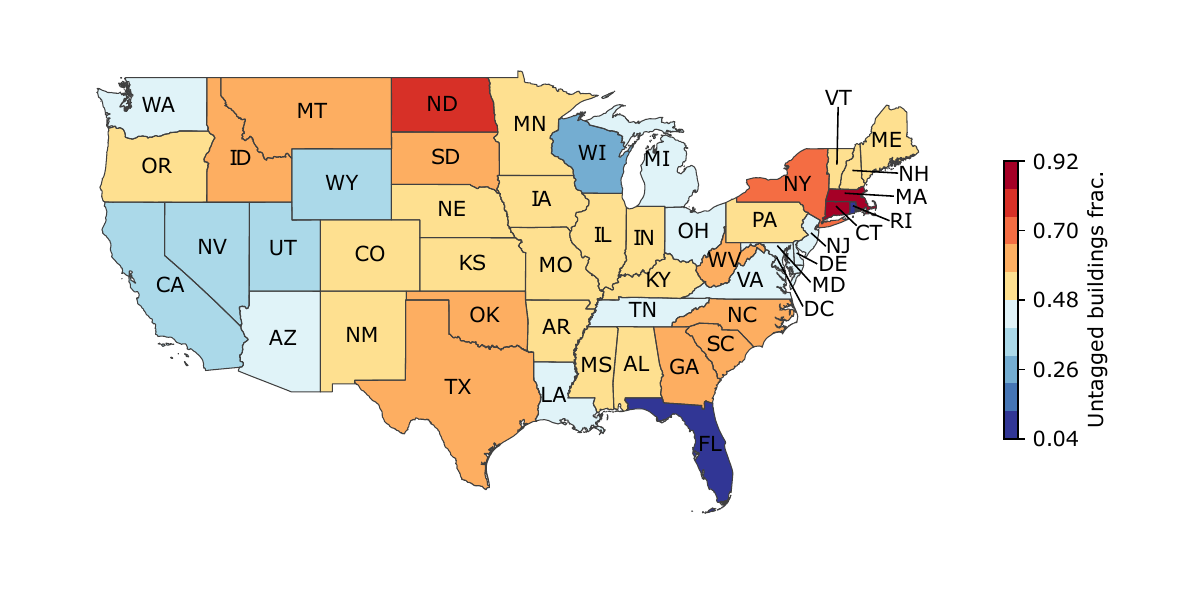}
    
    \caption{\textbf{Average proportion of buildings without annotations.} We considered both the tags associated with the buildings and the OSM geospatial features that overlap with the building footprints, such as ``landuse'' and ``amenity''. The states with the lowest fractions of untagged buildings are Rhode Island (0.04), Florida (0.11), Wisconsin (0.24), Washington D.C. (0.24), and Wyoming (0.34). The states with the highest proportions of untagged buildings are Massachusetts (0.92), Connecticut (0.85), North Dakota (0.76), New York (0.69), and West Virginia (0.64).}
    \label{fig:map_fraction}
\end{figure*}

Figure~\ref{fig:map_fraction} summarizes the state of annotation completeness in OSM data in the U.S. (as of August 2024). This map illustrates the average \textit{proportion of untagged buildings} across states in the contiguous U.S. 
In Figure~\ref{fig:map_fraction}, Rhode Island and Florida have the best coverage, while Massachusetts and Connecticut have the least.

Given incomplete annotations, efforts have been made to classify OSM building footprints. Fan \emph{et al.}~\cite{fan2014estimation} classifies building footprints using urban morphology and other features extracted from building footprints~\cite{fan2014estimation}. Bandam~\emph{et al.}~\cite{bandam2022classification} proposed a data-driven approach that combines external sources with OSM data (e.g., building heights). Atwal~\emph{et al.}~\cite{atwal2022predicting} proposed a machine-learning approach using both OSM-associated tags and features extracted from the footprints. The disadvantage of these approaches is that they require official administrative building footprints as training data. This work proposes an \textit{un}supervised method to identify whether a building is residential or non-residential based solely on the information captured in OSM, including building footprints and associated tags, and auxiliary data from features overlapping the footprint geometries, such as POIs and land use. Using this methodology, we create a comprehensive dataset that includes all buildings in the U.S. classified as residential and non-residential.

To validate our approach, we compare our results to a select set of official data. We use several counties from a metropolitan statistical area, namely \emph{Minneapolis and St. Paul}. We test the same approach using other regions in the U.S. to understand if the classifications are consistent. We further evaluate it to understand when our method is expected to perform better and to enhance comprehension of the results in arbitrary areas where we do not have ground truth.

\section*{Methods}
The main goal of this work is to provide a \textit{comprehensive U.S. building footprint dataset} with all buildings classified as residential or non-residential. We use an unsupervised method for this based on OSM building footprint information and their tags, as well as auxiliary data from other OSM features that can be inherited by the building footprints based on their spatial overlap. This section describes the data sources and the steps to achieve this.

The primary source of data to create our dataset is
OSM~\cite{OpenStreetMap}, a collaborative project to create a free and editable map of the world. OSM allows users to view, edit, and use geographic data in a collaborative way. For example, users can add data such as roads, trails, amenities, train stations, among others. The data is freely available under the \emph{Open Database License} (ODbL)~\cite{osm_copyright}, allowing it to be used for any purpose. To access this data, we use OSMnx~\cite{boeing2024modeling}, a Python package for retrieving information from OSM. In addition to allowing the building footprints to be downloaded, OSMnx can be used to retrieve information on street networks, urban amenities, and POIs, among other geospatial features.

In OSM, \emph{nodes}, \emph{ways}, \emph{areas}, and \emph{relations} are key data elements used to model geographic features.  A node is a geolocated single point representing simple features such as a bank or parking lot entrance and serving as a building block for more complex structures. A way is an ordered list of nodes representing features such as roads or rivers. An area (or filled polygon) is a special type of way that forms a closed loop and is used to represent geographic objects with a defined area, such as parks or building footprints. In OSM, any closed way can be interpreted as an area if the feature it represents is an enclosed space. A relation defines relationships between multiple nodes, ways, or other relationships. It is used for complex features, such as public transportation routes or multi-part boundaries, where simple ways or nodes are insufficient.

In OSM, data is organized using key-value pairs called tags. Tags describe the attributes of geographic features, where the key represents the type of attribute and the value specifies the details of the attribute. For instance, the \emph{building} key can have the value ``residential'', and the pair \emph{building} ``residential'' is a tag of the building footprint.

We also use official data delineating the boundaries of regions and sub-regions of the country to extract OSM features such as building footprints based on their belonging to a certain region. For this, we use the official boundaries of the \emph{counties}~\cite{us_boundary_files_2023}. Specifically, we use the \emph{1:500,000 (national)} file. This is a shapefile containing all counties or equivalent regions. We use the \emph{Annual Resident Population Estimates and Estimated Components of Resident Population Change for Metropolitan and Micropolitan Statistical Areas and Their Geographic Components for the United States}~\cite{us_census_bureau_2023} to determine whether these regions are metropolitan (a core area with a population of 50,000+)~\cite{glossary_census}, micropolitan (a core area with a population of >10,000 and <50,000)~\cite{glossary_census}, or other. The resulting classified building footprints obtained from this study are therefore organized by (i) metropolitan statistical area, (ii) micropolitan statistical area, and (iii) other.

Our building footprint classification pipeline relies on two different types of data: (i) the building footprints with their respective tags and (ii) auxiliary data. Here, we consider polygons with a \emph{building} key and any value as building footprints. Additionally, auxiliary data supplements the information beyond the building footprint tags. More specifically, when other OSM features spatially intersect with a building footprint, the building footprint inherits the tags associated with those features.
For example, when a building footprint intersects with a polygon with \emph{landuse} ``residential'' tag, the building footprint inherits this \emph{landuse} ``residential'' tag. Figure~\ref{fig:pipeline} shows a visual representation of our methodology, where the left panel shows the input data. The right side of the figure summarizes the processing pipeline, starting with the information associated with the building footprints. Next, we combine auxiliary data with the footprints to classify the unclassified buildings. 

Furthermore, Algorithm~\ref{alg:method} describes the proposed method for classifying building footprints in a given geographic area. The algorithm takes as input a polygon defining the geographic area of interest and several sets and dictionaries representing different categories of buildings, keys, and tags. Specifically, the inputs are as follows: the set of keys $\mathcal{S}_{\text{add}}$ contains the OSM keys used to download, retrieving all data elements that contain at least one of these keys (see the list in Supplementary Information~S1.1); the set of accommodation buildings $\mathcal{S}_{\text{acc}}$ (Supplementary Information~S1.2), which contains tags identifying accommodation-related buildings; the set of additional footprint keys $\mathcal{S}_{\text{add}}$, which contains keys associated with building footprints that may aid in classification (this is the same set of keys used to download the data from Supplementary Information~S1.1); the dictionary of tags to be skipped $\mathcal{D}_{\text{skip}}$ (Supplementary Information~S1.4.1), where the dictionary keys are OSM keys and the values are lists of values to be skipped during classification; the set of skipped values $\mathcal{S}_{\text{skip}}$ (Supplementary Information~S1.4.1), which contains tags to be ignored during classification; the dictionary of residential buildings $\mathcal{D}_{\text{res}}$ (Supplementary Information~S1.4.2), where the dictionary keys are the OSM keys and the values are the OSM values; the non-residential buildings dictionary $\mathcal{D}_{\text{non-res}}$ (Supplementary Information~S1.4.3); and the other non-residential auxiliary key set $\mathcal{S}_{\text{other\_non-res}}$ (Supplementary Information~S1.4.3), which contains additional tags used to identify non-residential buildings. The classification is performed by iterating over all the buildings in the area and applying the rules and conditions defined in the algorithm.

The entire pipeline is described as follows:

\begin{itemize}
    \item \textbf{Download OSM data:} First, for each county, we downloaded the data using OSMnx, specifying its official boundaries. To download additional geospatial features that are not directly associated with building footprints in OSM, we use a manually selected list of tags. We refer to the geospatial features that can be used to supplement the building footprint tags based on their spatial intersection as auxiliary data (see line 9 in Algorithm~\ref{alg:method}). See the full list of tags in Supplemental Information~S1.1;
    \item \textbf{Create auxiliary data:} To classify buildings when the building tag value is unknown, we use auxiliary data. Specifically, building footprints that spatially intersect with other OSM features can inherit the tags from these features.
    \item \textbf{Residential classification I:} All buildings with a building tag value related to a residential classification are classified as residential, except those with the value ``hotel'' ($\mathcal{S}_{\text{acc}}$). See the complete list of the tag values that we consider residential in Supplementary Information~S.1.2. Since sheds and garages can be represented as separate building footprints but are expected to be part of a house, these structures are also classified as residential. See lines 17--18 in Algorithm~\ref{alg:method};
    \item \textbf{Non-Residential classification I:} The footprints not classified as residential in the previous steps and have at least one building tag value, except ``service'', ``roof'', ``ruins'', and ``construction'', are classified as non-residential (see lines 19--20 in Algorithm~\ref{alg:method}).
    Building footprints can also have keys other than the building key-value pair ($\mathcal{S}_{\text{add}}$). All buildings not classified in the previous steps that have values in other keys (e.g., amenity key) are then classified as non-residential (see lines 22--27 in Algorithm~\ref{alg:method}). See Supplementary Information~S1.1;
    \item \textbf{Residential/Non-Residential classification II:} 
    Next, we consider the tags that the building footprints inherit from overlapping features. We consider a generic list of tags and tag values to be ignored ($\mathcal{D}_{\text{skip}}$ and $\mathcal{S}_{\text{skip}}$). If a tag or tag value appears on the list, it is excluded from consideration (See Supplementary Information S1.4.1.). For example, if a building footprint intersects with an area \emph{landuse} ``construction'' tag, this information is ignored.
    We also consider tags inherited from auxiliary data that are relevant to classifying the building footprint as residential ($\mathcal{D}_{\text{res}}$). For example, if a building intersects with a polygon with a \emph{landuse} ``residential'' tag, the building is classified as residential. Next, we consider tags from auxiliary data relevant for classifying buildings as non-residential ($\mathcal{D}_{\text{non-res}}$). For example, buildings that intersect with POI features with an office key will be classified as non-residential, regardless of the value of the key. In some cases, it is the tag values that are relevant for classifying non-residential buildings ($\mathcal{S}_{\text{other\_non-res}}$). For example, if a building overlaps with a POI with an \emph{amenity} tag value ``restaurant'', it is classified as non-residential, but not if the tag value is unclear, e.g., ``toilets''. See lines 28--41 in Algorithm~\ref{alg:method}. The lists of auxiliary data tags and values used are shown in Supplementary Information~S1.4.2~and~S1.4.3. 
    Note that we manually select the auxiliary data based on the OSM documentation;
    \item \textbf{Residential classification III:} Since most buildings are expected to be residential and the OSM users tend to add more information to the POIs, the remaining unknown buildings are classified as residential  (see lines 42--43 in Algorithm~\ref{alg:method}).
\end{itemize}

To alleviate potential memory and computation problems, we divide the region into rectangular sub-regions, download the data, and merge the resulting buildings. Here, we take this parameter as $5 \times 5$ rectangles that cover the entire area of interest and download the data from the rectangles that overlap with the area of interest. Since the buildings on the boundaries may be downloaded more than once, we remove the duplicate information after merging the rectangular sub-regions.

\section*{Data Records}
We organized the dataset files according to three categories, namely Metropolitan Statistical Areas, Micropolitan Statistical Areas, and the remaining counties, which are located in the \emph{metropolitan}, \emph{micropolitan}, and \emph{other} folders. These categories were given by the ``Annual Resident Population Estimates and Estimated Components of Resident Population Change for Metropolitan and Micropolitan Statistical Areas and Their Geographic Components for the United States''. 

The regions are named according to their respective Core-Based Statistical Area (CBSA) codes to link the files to the official data. The names of the county shape files follow the naming standard \emph{``STCOU\_county name.shp''}, in which STCOU is the State-County code of a particular region. For example, the file for Fairfax County, VA, is in the path \emph{metropolitan/47900/Fairfax\_51059.shp}, where 47900 is the CBSA and 51059 is the STCOU. 
STCOU is also known as GEOID or FIPS (Federal Information Processing System), where the first two digits represent the state-level FIPS code, and the last three digits represent the county FIPS code. In this example, 51 is the state code, and 059 is the county FIPS code.
We define the projection of the files according to the Universal Transverse Mercator (UTM) projection, which better matches the center of mass of the country. For more details on defining the appropriate UTM, see Supplementary Information~S2. Note that the shapefiles can be projected in different UTM projections for the same metropolitan or micropolitan statistical area. Therefore, in an application where counties are merged, it is necessary to convert all files to the same coordinate. 

In addition to the geometry and the \emph{type} column representing the classification between residential and non-residential (``RES'' or ``NON\_RES''), the output shapefiles contain the following information. The \emph{tag used} column stores information about the tag, followed by its value, which is used to classify the building footprint. For example, if a building footprint contains the building tag with the value ``residential'', it is assigned to the \emph{tag used} column as ``building: residential''. Finally, the \emph{aux info} column is used to understand in which step of our method the building is classified. The different values of this column are (i) buildings that are classified as residential according to the building tag (``residential\_types''); (ii) buildings that are classified as non-residential according to the building tag (``non\_residential\_types''); (iii) for those buildings that are not classified due to lack of a building tag, but have another non-residential tag associated with them (``non\_residential\_aux\_tag''); (iv) for the auxiliary data steps, we first consider, in order of priority, the auxiliary data associated with residential buildings (``residential\_auxiliary''); (v) next, we consider the specific non-residential tags (``non\_residential\_auxiliary''); (vi) the non-specific auxiliary is considered (``non\_residential\_auxiliary\_generic\_tag''); and (vii) as a last step, the remaining building footprints are classified as residential (``residential\_unknown\_tag'').

The data is available in an OSF repository at \url{https://osf.io/utgae/}.

\section*{Technical Validation}
In order to validate the generated dataset, we compare the obtained results with the official data of different U.S. regions where building type data is available. In the following subsections, we present the data used for the validation and provide details on how we validate our building footprint classification. 

\subsection*{Validation Approach and Data} 
Ground truth datasets from official sources are typically available in two formats: either building footprints with the building type or detailed land use and zoning of the region type. The former offers a one-to-one comparison of the predicted building classification and the official classification. In the latter, we compare the predicted classification of the building footprints with the intersecting land use polygon. If the building overlaps with multiple land use polygons in the ground truth data, we assign the label corresponding to the area with the largest overlap. We note that we ignore all buildings characterized as mixed-use, as they do not fit into the binary classification of residential or non-residential. Furthermore, if a building does not overlap with the ground truth, it is excluded from the analysis.

We validate our unsupervised classification approach by comparing with ground truth data in two case studies: Minneapolis and St. Paul areas, where all counties belong to the same region, potentially introducing a regional bias in the quality of annotations. To address this, we include a second case study - a set of regions from different parts of the U.S. to demonstrate that our method generalizes beyond a single region, which includes the counties of Baltimore, MD; Hanover, VA; and Mecklenburg, NC; the city of Boulder, CO; and the city and county of Fairfax, VA. These additional regions were selected based on the availability of high-quality data. We compare the real building classification from these regions with our predicted classification and evaluate the performance of our method using Recall, Precision, and F1-Scores, considering each building footprint as a sample in the dataset. We compare the resulting building classification with the ground truth.

Here, we describe the validation datasets - administrative ground truth data containing information on residential and non-residential characteristics of the building and provide the references to download them.

\begin{itemize}
    \item \textbf{Minneapolis and St. Paul:} The ``Generalized Land Use Inventory'' dataset was created by the Metropolitan Council and includes Anoka, Carver, Dakota, Hennepin, Ramsey, Scott, and Washington counties in Minnesota. The dataset was derived from aerial imagery taken on April 4,5 and 10, 2020, and supplemented with county parcel and assessor data, online resources, field inspections, and community feedback~\cite{minneapolis2020generalized};
    \item \textbf{Baltimore, MD:} This dataset consists of the land use for parcels in Baltimore, MD County. Here, we use the data updated on September 19, 2023~\cite{baltimore2023landuse};
    \item \textbf{Boulder, CO:} For Boulder, we use the same data as for Atwal~\emph{et al.}~\cite{atwal2022predicting}. This is the official data for the city of Boulder, CO~\cite{boulder2022predicting};
    \item \textbf{Fairfax, VA:} We considered both the city and the county of Fairfax, VA. To do this, we renamed and merged the column representing land use to be the same in both datasets~\cite{fairfax_county2024landuse,fairfax_city2024landuse}. In the case of Fairfax County, we used the ``Existing Landuse generalized'' file, updated on April 13, 2024~\cite{fairfax_city2024landuse};
    \item \textbf{Hanover, VA:}  We use the official Hanover County, VA data of the ``Zoning Districts''~\cite{hanover2024zoning}.
    \item \textbf{Mecklenburg, NC:} We use the official data from Mecklenburg County, NC, namely ``Tax Parcel Landuse Existing''
~\cite{mecklenburg2024landuse}.
\end{itemize}

The conversion between the official data for residential (RES), non-residential (NON\_RES), and unaccounted (N/A) buildings is shown in Supplementary Material~S3.

\subsubsection*{Case Study I: Minneapolis and St. Paul}
For all areas considered in the validation, we downloaded the data via OSMnx using the convex hull of the considered ground truth to avoid problems due to possible differences between the ground truth and the official county boundaries. The results for the Minneapolis and St. Paul Metropolitan Council validations are summarized in Table~\ref{tab:msa_results}. As expected, the recall for residential buildings is high (close to 1) in all cases. However, this measure is significantly lower for non-residential buildings, where the worst recall, 0.59, was obtained for Dakota, MN. In contrast, all precision values for non-residential buildings are between 0.99 and 0.96, indicating that the proposed method provides relatively high precision for non-residential buildings. In addition, the F1-Scores do not vary significantly within the two classes and, as expected, are slightly higher for residential buildings.

For certain applications, the footprints of sheds and garages may not be useful. For instance, if the data is used to estimate population density, one could decide not to include sheds and garages. In general, our validation results improve only slightly when we remove sheds and garages. The exception is Dakota, MN, which improves by a lot with a recall of 0.59 when all buildings are included and a recall of 0.77 when sheds and garages are removed. The results where sheds and garages are excluded are found in Supplementary Information~S4.

\subsubsection*{Case Study II: Analysis Across Multiple Regions}
We expanded the testing to other cities and counties in the U.S. to check for consistency. The regions considered and the results are shown in Table~\ref{tab:us_regions}. Overall, in comparison to the Minneapolis and St. Paul Council, the performance is relatively the same. Among these other regions, the worst precision, 0.85, is found for the City of Boulder, CO. The results without considering the sheds and garages can be seen in Table~S2 of Supplementary Information~S4.

\begin{figure}[h]
    \centering
    \includegraphics[width=1\textwidth]{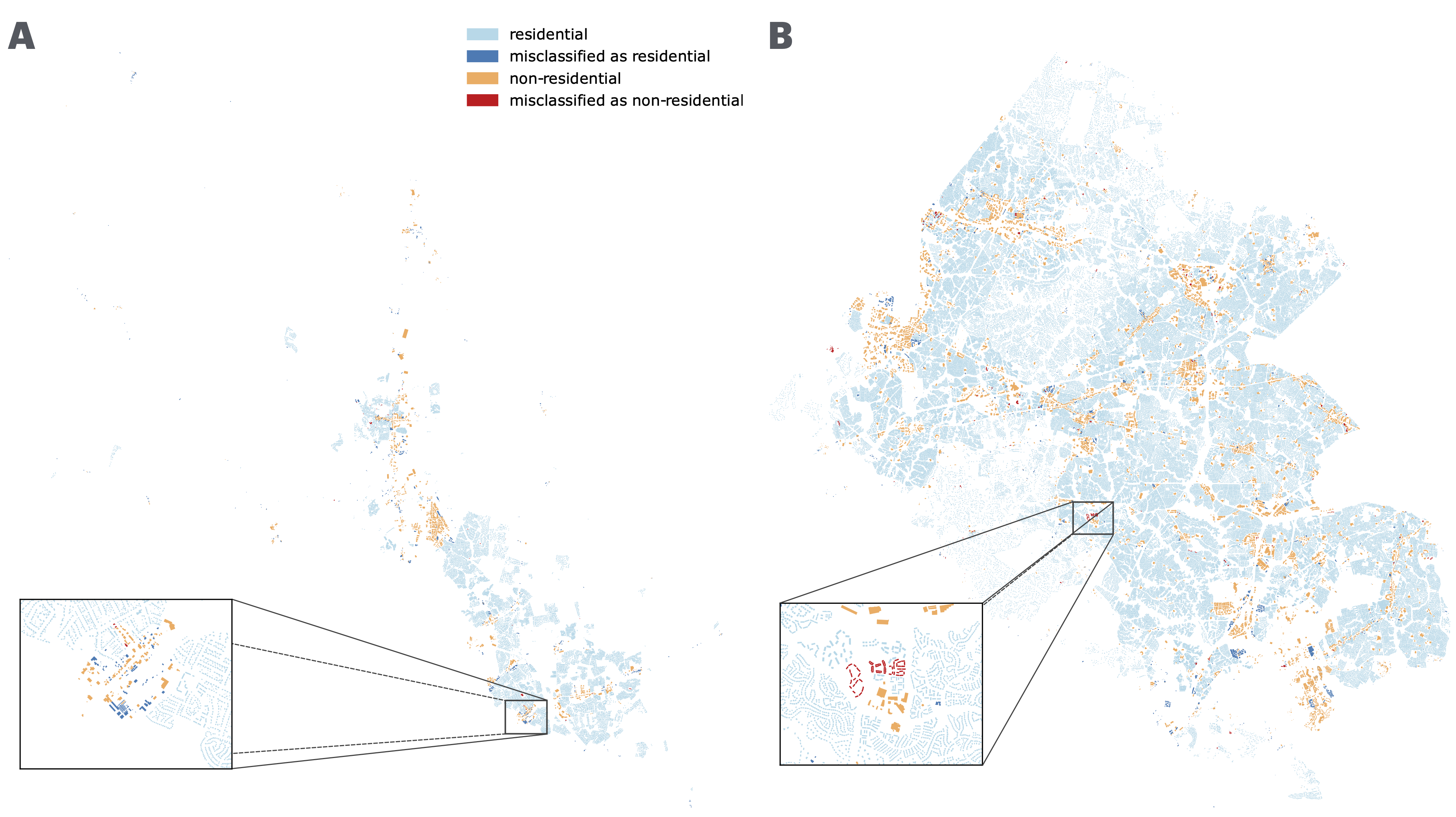}
    \caption{\textbf{Illustrations of the identified buildings.} Panel (a) shows Hanover, VA, and panel (b) shows Fairfax, VA. The mixed-use and unknown building footprints are not shown.}
    \label{fig:indentified_buildings}
\end{figure}

To better understand where our methodology is not working, we visualize the building footprints of two regions in Virginia, Hanover (Figure~\ref{fig:indentified_buildings}A) and Fairfax (Figure~\ref{fig:indentified_buildings}B), which have the worst and best average F1-Scores, respectively. In both cases, the misclassified buildings tend to be close to non-residential areas. The majority of the errors are non-residential buildings that have been misclassified as residential (see the dark blue buildings in the inset map in Figure~\ref{fig:indentified_buildings}A). However, in some cases, residential buildings are misclassified as non-residential (see the red buildings in the inset map in Figure~\ref{fig:indentified_buildings}B).

\subsubsection*{Investigating Causes of Building Misclassification}
\label{sec:outcome_evaluation}
\begin{figure*}[h]
    \centering
    \includegraphics[width=.85\textwidth]{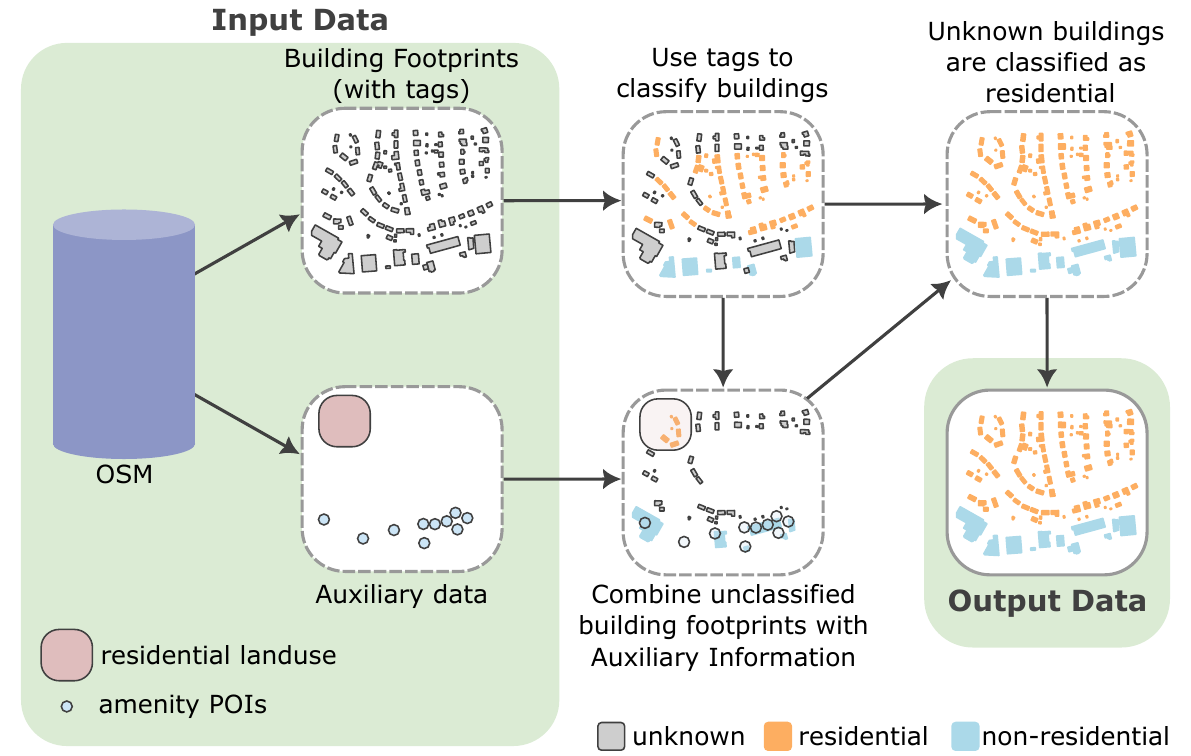}
    \caption{\textbf{Scheme of the building classification methodology.} The OSM data is obtained in two separate data types: the building footprints with their respective tag values and additional auxiliary data (e.g., land use and amenities). First, the buildings are classified with tags indicating residential and non-residential use. Next, the unknown buildings are classified using the additional auxiliary data that overlaps with the building footprints. Finally, the remaining unknown buildings are classified as residential.}
    \label{fig:pipeline}
\end{figure*}

Considering our approach, shown in Figure~\ref{fig:pipeline}, we classify buildings through a sequence of steps that examine footprint tags and additional auxiliary data obtained when footprints overlap with other geospatial features found in OSM data. As observed in the previous section, the most common errors are non-residential buildings that are misclassified as residential. In this section, we identify and analyze the reasons for these misclassification. 

\begin{figure}[h]
    \centering
    \includegraphics[width=0.5\textwidth]{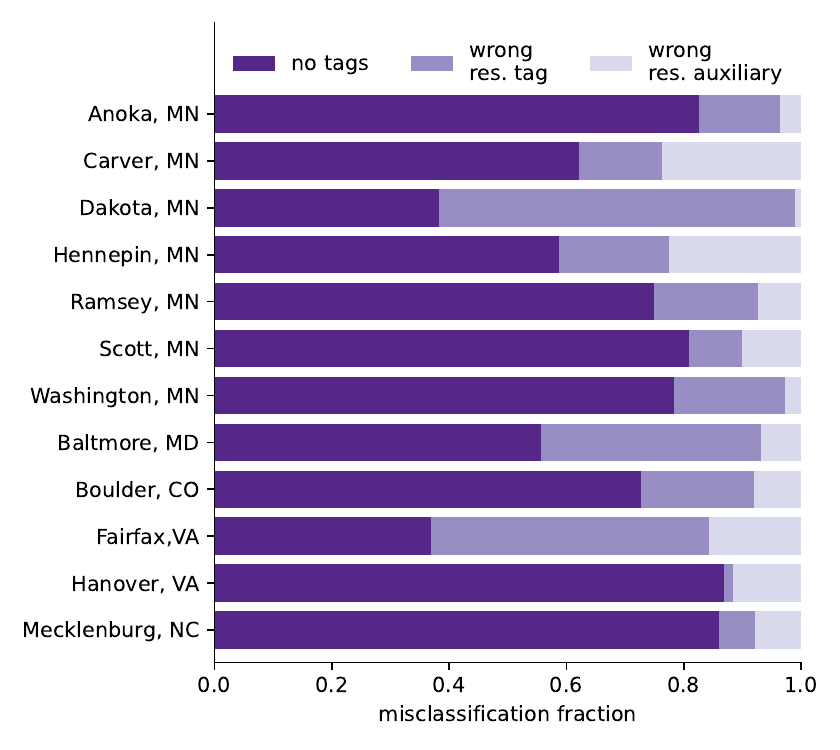}
    \caption{\textbf{Stacked bar chart showing the fractions of misclassified buildings.} The colors represent buildings misclassified due to the absence of tags (no tags), the presence of a residential tag (wrong res. tag), and incorrect residential auxiliary data (wrong res. auxiliary).}
    \label{fig:errors}
\end{figure}

The stacked bar chart in Figure~\ref{fig:errors} summarizes the reasons behind the misclassification of non-residential buildings as residential. In general, buildings are primarily misclassified as residential because they lack tags in OSM. A secondary reason for this misclassification is that some buildings were incorrectly tagged as residential in OSM. A third and less common source of error is incorrect auxiliary data. In almost all cases, the incorrect auxiliary data is the \emph{landuse} with a value of ``residential''. Dakota, MN, and Fairfax, VA, are the only exceptions to this order, where the most common error is not due to missing tags, but instead due to being incorrectly tagged as residential in OSM.  For example, of the buildings that were misclassified as residential in Fairfax, 47\% were tagged as residential in OSM, and 37\% had no tags at all. About 16\% of the misclassifications stemmed from combining the footprints with auxiliary data from other geospatial features, e.g., building footprints that overlapped with residential land use features. Interestingly, Fairfax, VA and Dakota, MN are the regions with the highest average F1 scores \ref{tab:us_regions}, suggesting that these regions may have been better annotated in OSM than the other regions analyzed.

\begin{figure}[h]
    \centering
    \includegraphics[width=0.5\textwidth]{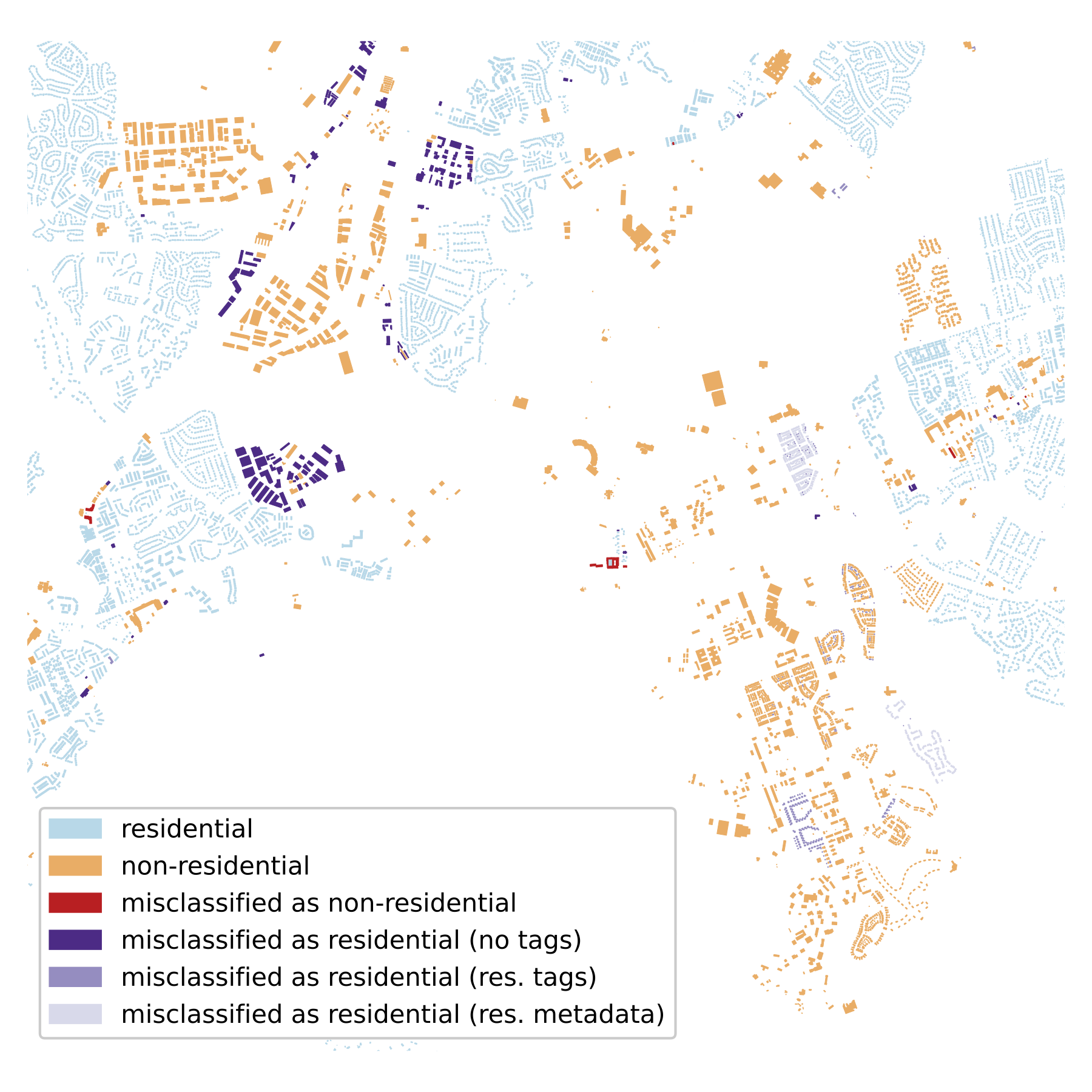}
    \caption{\textbf{Zoom in on a region of Fairfax.} This is a zoom in on Figure~\ref{fig:indentified_buildings}A, where we found buildings misclassified as residential.}
    \label{fig:missclassification}
\end{figure}

Figure~\ref{fig:missclassification} illustrates an example of buildings misclassified as residential within Fairfax neighborhoods (purple) and the reason for these misclassifications (different gradients of purple). 

\subsubsection*{National Analysis}
After validating our approach with official data, we execute it for all counties in the U.S. To download the data, we used the official polygons of each county. Only two counties returned no buildings: Wheeler, NE, and Rose Island, AS. 

\begin{figure*}[h]
    \centering
    \includegraphics[width=1.0\textwidth]{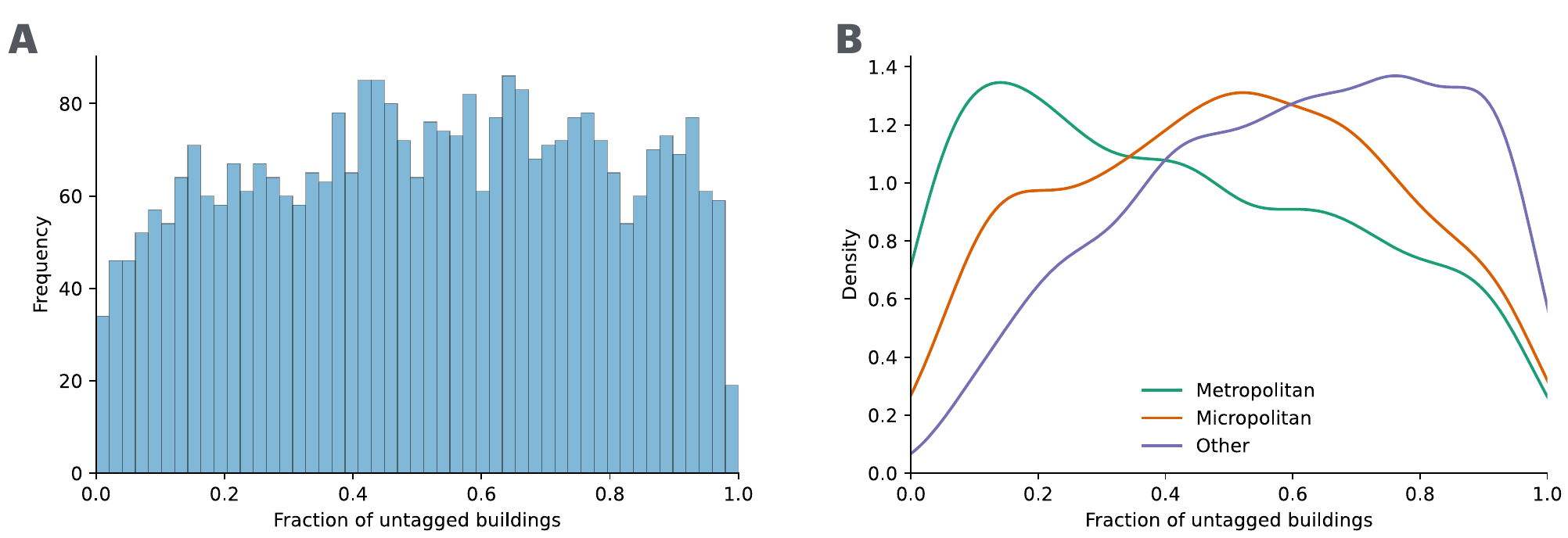}
    \caption{\textbf{Proportion of buildings without annotations.} Panel (a) shows the histogram of the ratios for all counties in the entire U.S. Panel (b) shows the comparison of the density of the distributions for metropolitan, micropolitan, and other areas.}
    \label{fig:fraction}
\end{figure*}

In the previous section, we show that one limitation of OSM is the lack of annotated data. To understand the extent to which this affects the obtained dataset, we further analyze it to determine how the regions are annotated. Here, we consider the annotation of the buildings in combination with the auxiliary data, which is summarized in Figure~\ref{fig:map_fraction}. Note that the auxiliary data considered here is limited to those we consider in our method (described in Methods). Figure~\ref{fig:fraction}A shows the fraction of annotated buildings in the US for all counties. The average fraction of annotated buildings per county is 0.51. The least annotated areas are Emmons County, ND; Monroe County, MO; Gem County, ID; Throckmorton County, TX; and Northern Islands Municipality, MP. All of these are regions with a low number of buildings, and less than 2\% of buildings are annotated. In contrast, the most annotated areas are Carlisle County, KY; Falls Church City, VA; Nassau County, NY; Arlington County, VA; and Fairfax City, VA. Arlington County, VA, is a medium county, and Nassau County, NY, is a large county. The other regions are relatively small. For a graphical representation of the U.S. map with fractions of untagged buildings, see Figure~S2 in Supplementary Material~S5.

There is no evidence that the total number of annotations affects the quality of the classification when comparing the regions analyzed so far. The recall for non-residential buildings in Dakota, MN, is 0.59, while the proportion of buildings without annotations is 0.03. In contrast, Ramsey, MN, has a similar recall (0.61) with a proportion of unannotated buildings of 0.2. Furthermore, the least annotated county tested is Hanover, VA, with a fraction of 0.59 and a recall of 0.62. Overall, the recall values for nonresidential buildings presented in the previous sections indicate that not all nonresidential buildings are tagged. This suggests that the fraction of tags is not sufficient to determine whether the residential/non-residential classification will be of good quality.

We also compare the proportions of untagged buildings for different types of regions, namely metropolitan, micropolitan, and other areas (see Figure~\ref{fig:fraction}B). As can be seen, metropolitan regions tend to be more annotated than micropolitan regions. Counties in the ``other'' category (neither metropolitan nor micropolitan regions) tend to have the fewest annotations. The average fractions of unannotated buildings are 0.42, 0.50, and 0.60 for metropolitan, micropolitan, and other, respectively. 

\section*{Usage Notes}
We have projected the data to UTM to facilitate applications where distances are important. However, for applications where it is necessary to merge counties, the user needs to convert them all to the same coordinate. Furthermore, if one wants to use the Coordinate Reference System (CRS), this can easily be done using Python GeoPandas or GIS (Geographic Information System) software. 

Buildings that straddle the boundaries of two different counties are part of more than one shapefile. Therefore, when merging files, it is necessary to eliminate duplicate information. 

If one wants to use our dataset to avoid sheds and garages, it can be easily filtered by using the \emph{`tag used'} column. See the code example in Figure~S3 of Supplementary Information~S6.

\section*{Code availability}

The code is available in a GitHub repository at \url{https://github.com/gmuggs/OSM-Building-Classification}.

\bibliography{refs}

\section*{Author contributions statement}
H.F.A, S.M.R., and D.P. conceptualization and methodology; H.F.A, software, validation, formal analysis, visualization, and writing - original draft; H.F.A, S.M.R., S.R., K.S.A, H.K., T.A., and D.P. investigation; H.F.A and S.R. data curation; H.K., T.A., and D.P. resources, project administration, and funding acquisition; All authors reviewed the manuscript.

\section*{Competing interests}
The author declares no competing interests.

\section*{Figures \& Tables}

\begin{algorithm}[!h]
  \caption{Classification of all the building footprints in a given geographical area.}
   \label{alg:method}
   \begin{algorithmic}[1]
   \State \textbf{Input:} 
   \State Polygon of the area \VAR{polygon}; Accommodation buildings set $\mathcal{S}_{\text{acc}}$; 
   \State Additional footprint tags set $\mathcal{S}_{\text{add}}$; 
   \State Dictionary with tags as keys and list of values to be skipped as values $\mathcal{D}_{\text{skip}}$; 
   \State Skipping tags set $\mathcal{S}_{\text{skip}}$; 
   \State Dictionary with tags as keys and residential buildings as values $\mathcal{D}_{\text{res}}$; 
   \State Dictionary with tags as keys and non-residential buildings as values $\mathcal{D}_{\text{non-res}}$; 
   \State Other non-residential auxiliary tags $\mathcal{S}_{\text{other\_non-res}}$;
   
   \State \textbf{Output:} 
   \State Resultant geodataframe \VAR{out} with building classified as residential or non-residential; 
   
   \State \VAR{gdf} $\gets$ All OSM data enclosed by \VAR{polygon};
   \State \VAR{buildings} $\gets$ all buildings of \VAR{gdf};
   \State \VAR{auxiliary} $\gets$ \VAR{gdf} data except the entries of \VAR{buildings};
   \State \VAR{out} $\gets$ geodataframe with all buildings and \VAR{'type'} column set to \VAR{null};
   \State \VAR{unknown} $\gets$ \VAR{\{'yes', 'service', 'roof', 'ruins', 'construction'\}};
   \For{\VAR{building} $\in$ \VAR{buildings}}
        \If{\VAR{building['building']} $\in$ $\mathcal{S}_{\text{acc}}$} \Comment{Using the information associated with the building footprint}
            \State \VAR{out['type']} $\gets$ \VAR{'RES'};
        \ElsIf{\VAR{building['building']} $\notin$ \VAR{unknown}} \Comment{The value `yes' means an unspecified value} 
            \State \VAR{out['type']} $\gets$ \VAR{'NON\_RES'};
        \Else \Comment{Use other tags associated with the building footprint}
            \For{\VAR{tag} $\in$ $\mathcal{S}_{\text{add}}$}
                \If{\VAR{building[tag]} $!=$ \VAR{null}}
                    \State \VAR{out['type']} $\gets$ \VAR{'NON\_RES'};
                \EndIf
            \EndFor
        \EndIf 
        \If{\VAR{out['type']} $==$ \VAR{null}} \Comment{If not classified with the previous information, use the auxiliary data}
            \State \VAR{inter\_building} $\gets$ \VAR{building} spatially intersected with \VAR{auxiliary};

            \For{\VAR{key} $\in$ keys($\mathcal{D}_{\text{skip}}$)}
                \If{\VAR{key} in \VAR{inter\_building} \AND \VAR{inter\_building[key]} in $\mathcal{D}_{\text{skip}}$\VAR{[key]}}
                    \State Drop \VAR{key} from \VAR{inter\_building};
                \EndIf
            \EndFor
            
            \State \VAR{inter\_building} $\gets$ \VAR{inter\_building} $\setminus$ $\mathcal{S}_{\text{skip}}$;
            \If{exists \VAR{tag} $\in$ \VAR{inter\_building} s.t. \VAR{tag} $\in$ \{(\VAR{key}, \VAR{val}) $\mid$ \VAR{key} $\in$ $\mathcal{D}_{\text{res}} \text{ and }$ \VAR{val} $\in \mathcal{D}_{\text{res}}$\VAR{[key]}\}}
                \State \VAR{out['type']} $\gets$ \VAR{'RES'};
            \ElsIf{exists \VAR{tag} $\in$ \VAR{inter\_building} s.t. \VAR{tag} $\in$ \{(\VAR{key}, \VAR{val}) $\mid$ \VAR{key} $\in$ $\mathcal{D}_{\text{non-res}} \text{ and }$ \VAR{val} $\in \mathcal{D}_{\text{non-res}}$\VAR{[key]}\}}
                \State \VAR{out['type']} $\gets$ \VAR{'NON\_RES'};
            
            \ElsIf{exists \VAR{x} $\in$ \VAR{inter\_building} s.t. \VAR{x} $\in$ $\mathcal{S}_{\text{other\_non-res}}$}

                \State \VAR{out['type']} $\gets$ \VAR{'NON\_RES'};
            \Else
                \State \VAR{out['type']} $\gets$ \VAR{'RES'};
            \EndIf
        \EndIf                    
   \EndFor
   \end{algorithmic}
\end{algorithm}

\begin{table*}[h]
\centering
\begin{tabular}{|c|c|c|c|c|c|}
\hline
\textbf{County}             &\textbf{Class} &\textbf{Precision}&\textbf{Recall}&\textbf{F1-Score}&\textbf{Avg. F1-Score} \\ \hline \hline
\multirow{2}{*}{Anoka, MN}  &non-residential&   0.99    &  0.69  &   0.81   &  \multirow{2}{*}{0.88} \\ \cline{2-5} 
                            &  residential  &   0.90    &  1.00  &   0.95   &  \\ \hline
\multirow{2}{*}{Carver, MN} &non-residential&   0.96    &  0.74  &   0.84   &  \multirow{2}{*}{0.91} \\ \cline{2-5}
                            &  residential  &   0.96    &  1.00  &   0.98   &     \\ \hline
\multirow{2}{*}{Dakota, MN} &non-residential&   0.98    &  0.59  &   0.73   &  \multirow{2}{*}{0.86} \\ \cline{2-5} 
                            &  residential  &   0.97    &  1.00  &   0.98   &     \\ \hline
\multirow{2}{*}{Hennepin, MN} &non-residential& 0.97    &  0.75  &   0.85   &  \multirow{2}{*}{0.92} \\ \cline{2-5} 
                            &  residential  &   0.97    &  1.00  &   0.99   &     \\ \hline
\multirow{2}{*}{Ramsey, MN}  &non-residential&   0.97    &  0.61  &   0.75   &  \multirow{2}{*}{0.86} \\ \cline{2-5} 
                            &  residential  &   0.94    &  1.00  &   0.97   &     \\ \hline
\multirow{2}{*}{Scott, MN}   &non-residential&   0.98    &  0.69  &   0.81   &  \multirow{2}{*}{0.89} \\ \cline{2-5} 
                            &  residential  &   0.95    &  1.00  &   0.97   &     \\ \hline
\multirow{2}{*}{Washington, MN} &non-residential&0.98   &  0.73  &   0.83   &  \multirow{2}{*}{0.91} \\ \cline{2-5} 
                            &  residential  &   0.96    &  1.00  &   0.98   &     \\ \hline
\end{tabular}
\caption{\textbf{Prediction results for the Minneapolis and St. Paul area.} Results for the Minneapolis and St. Paul metropolitan area counties.}
\label{tab:msa_results}
\end{table*}

\begin{table*}[h]
\centering
\begin{tabular}{|c|c|c|c|c|c|}
\hline
\textbf{Region}                     &\textbf{Class} &\textbf{Precision}&\textbf{Recall}&\textbf{F1-Score}&\textbf{Avg. F1-Score} \\ \hline \hline
\multirow{2}{*}{Baltimore, MD}  &non-residential&   0.94    &  0.81  &   0.87   &  \multirow{2}{*}{0.93} \\ \cline{2-5} 
                                    &  residential  &   0.99    &  1.00  &   0.99   &  \\ \hline 
\multirow{2}{*}{Boulder, CO$^{*}$} &non-residential&   0.85    &  0.70  &   0.77   &  \multirow{2}{*}{0.88} \\ \cline{2-5} 
                                    &  residential  &   0.97    &  0.99  &   0.98   &  \\ \hline 
\multirow{2}{*}{Fairfax, VA$^{**}$}    &non-residential&   0.95    &  0.78  &   0.86   &  \multirow{2}{*}{0.93} \\ \cline{2-5} 
                                    &  residential  &   0.99    &  1.00  &   0.99   &  \\ \hline
                                    
\multirow{2}{*}{Hanover, VA}    &non-residential&   0.97    &  0.62  &   0.75   &  \multirow{2}{*}{0.87} \\ \cline{2-5} 
                                    &  residential  &   0.97    &  1.00  &   0.98   &  \\ \hline 
                                    
\multirow{2}{*}{Mecklenburg, NC}&non-residential&  0.92  &  0.74  &   0.82   &  \multirow{2}{*}{0.91} \\ \cline{2-5} 
                                    &  residential  &   0.98    &  1.00  &   0.99   &  \\ \hline 
\end{tabular}
\caption{\textbf{Prediction results for other regions of the United States.} Here, Boulder$^*$ is a city, Fairfax$^{**}$ is both the county and the town of Fairfax, and the others are counties.}
\label{tab:us_regions}
\end{table*}

\clearpage
\newpage

\section*{Supplementary Information}

\thispagestyle{empty}

\renewcommand{\thefigure}{S\arabic{figure}}
\setcounter{figure}{0}

\renewcommand{\thetable}{S\arabic{table}}
\setcounter{table}{0}

\renewcommand{\thesection}{S\arabic{section}}
\setcounter{section}{0}

\section{Selected auxiliary information}
\label{sup:tags}
In this section, we divide the building tags into residential and non-residential categories. In addition, we show the tags of the auxiliary information used for those buildings that are not classified by the building tags.

\subsection{Download data}
\label{sup:download_tags}
In order to download data using OSMnx, we must specify which keys we have used. The set of keys used here is as follows. In OSMnx, one of the parameters of the method used to download data (\emph{features\_from\_polygon}) is called \emph{tag}, which specifies the OSM keys. This parameter allows the method to download all data elements that contain tags with the specified keys.
The key \emph{surface} is only used to reduce errors caused by trying to download empty data and is removed for other analyses. 

\begin{tasks}(4)
    \task[$\bullet$] building; 
    \task[$\bullet$] surface;
    \task[$\bullet$] amenity;
    \task[$\bullet$] emergency;
    \task[$\bullet$] healthcare;
    \task[$\bullet$] landuse;
    \task[$\bullet$] military;
    \task[$\bullet$] office;
    \task[$\bullet$] public\_transport;
    \task[$\bullet$] service;
    \task[$\bullet$] shop;
    \task[$\bullet$] sport;
    \task[$\bullet$] telecom;
    \task[$\bullet$] tourism;
    \task[$\bullet$] brand;
    \task[$\bullet$] clothes;
    \task[$\bullet$] leisure;
    \task[$\bullet$] cemetery.
\end{tasks}

\subsection{Accomodation tags}
\label{sup:tags_res}
Here, we list the building values for the \emph{building} key, indicating that the tag is residential. 

\begin{tasks}(4)
    \task[$\bullet$] apartments;
    \task[$\bullet$] barracks;
    \task[$\bullet$] bungalow;
    \task[$\bullet$] cabin;
    \task[$\bullet$] detached;
    \task[$\bullet$] dormitory;
    \task[$\bullet$] farm;
    \task[$\bullet$] ger;
    \task[$\bullet$] house;
    \task[$\bullet$] houseboat;
    \task[$\bullet$] residential;
    \task[$\bullet$] semidetached\_house;
    \task[$\bullet$] static\_caravan;
    \task[$\bullet$] stilt\_house;
    \task[$\bullet$] terrace;
    \task[$\bullet$] tree\_house;
    \task[$\bullet$] trullo;
    \task[$\bullet$] townhouse;
    \task[$\bullet$] townhome;
    \task[$\bullet$] boathouse;
    \task[$\bullet$] shed;
    \task[$\bullet$] garage; 
    \task[$\bullet$] garages.
\end{tasks}

\subsection{Non-residential tags}
\label{sup:tags_other_tags}
The set of non-residential tags consists of all other building tag values not in Section~\ref{sup:tags_res}, except ``yes''. The ``yes'' value represents that the building is of an unknown type. 

\subsection{Selected auxiliary information lists}
\label{sup:tags_metadata}
This set is divided into two sublists, one for residential and one for non-residential buildings. 

\subsubsection{Skipped tags}
\label{sup:skip_tags_metadata}
Our method ignores the following tags. 
For the key \emph{landuse}, if the value is ``forest'' the tag is skipped, and for the key ``leisure'', if the value is ``park'' or ``swimming\_pool'' the tag is skipped. The following set of values are skipped for all keys:
\begin{tasks}(4)
    \task[$\bullet$] construction;
    \task[$\bullet$] driveway;
    \task[$\bullet$] grass;
    \task[$\bullet$] farmyard;
    \task[$\bullet$] farmland;
    \task[$\bullet$] nature\_reserve.
\end{tasks}

\subsubsection{Residential buildings}
\label{sup:res_metadata}
For the key \emph{landuse}, we only consider the value \emph{residential}. Next, for the key \emph{tourism}, we consider \emph{apartment} and \emph{guest\_house}. Note that our method considers the key \emph{landuse} first.

\subsubsection{non-residential buildings}
\label{sup:non_res_metadata}
The non-residential keys and their values are listed below. The values are considered in the order in which they appear in the list. 

$\bullet$ \textbf{landuse}:
\begin{tasks}(4)
  \task[\quad --] \ commercial; 
  \task[\quad --] \ retail;
  \task[\quad --] \ industrial;
  \task[\quad --] \ institutional;
  \task[\quad --] \ education;
  \task[\quad --] \ military; 
  \task[\quad --] \ port;
  \task[\quad --] \ religious;
  \task[\quad --] \ winter\_sports;
  \task[\quad --] \ cemetery;
  \task[\quad --] \ grave\_yard.
\end{tasks}

$\bullet$ \textbf{amenity}:
\begin{tasks}(4)
  \task[\quad --] \ courthouse;
  \task[\quad --] \ fire\_station;
  \task[\quad --] \ police;
  \task[\quad --] \ post\_depot;
  \task[\quad --] \ post\_office;
  \task[\quad --] \ prison;
  \task[\quad --] \ ranger\_station;
  \task[\quad --] \ townhall;
  \task[\quad --] \ college;
  \task[\quad --] \ kindergarten;
  \task[\quad --] \ library;
  \task[\quad --] \ research\_institute;
  \task[\quad --] \ school;
  \task[\quad --] \ university;
  \task[\quad --] \ car\_rental;
  \task[\quad --] \ car\_wash;
  \task[\quad --] \ vehicle\_inspection;
  \task[\quad --] \ ferry\_terminal;
  \task[\quad --] \ fuel;
  \task[\quad --] \ hospital;
  \task[\quad --] \ brothel;
  \task[\quad --] \ casino;
  \task[\quad --] \ cinema;
  \task[\quad --] \ conference\_centre;
  \task[\quad --] \ events\_venue;
  \task[\quad --] \ exhibition\_centre;
  \task[\quad --] \ love\_hotel;
  \task[\quad --] \ nightclub;
  \task[\quad --] \ planetarium;
  \task[\quad --] \ theatre;
  \task[\quad --] \ bar;
  \task[\quad --] \ restaurant.
\end{tasks}

The buildings that are not classified with the previous tags and that contain a non-null value with the following keys are considered non-residential. 

\begin{tasks}(4)
    \task[$\bullet$] emergency;
    \task[$\bullet$] healthcare;
    \task[$\bullet$] landuse;
    \task[$\bullet$] military;
    \task[$\bullet$] office;
    \task[$\bullet$] public\_transport;
    \task[$\bullet$] service;
    \task[$\bullet$] shopv;
    \task[$\bullet$] sport;
    \task[$\bullet$] telecom;
    \task[$\bullet$] tourism;
    \task[$\bullet$] brand;
    \task[$\bullet$] clothes;
    \task[$\bullet$] leisure;
    \task[$\bullet$] cemetery.
\end{tasks}

\section{Defining UTM}
\label{sup:convert_coords}
All the data generated from this study is projected using the Universal Transverse Mercator (UTM) Coordinate Reference System (CRS). To find the best coordinates to project, we used the Python function shown in Figure~\ref{fig:code_coord}, in which the input is a \emph{GeoDataFrame} of \emph{Geopandas}. For more information, see Section 6.3 of ref.~\cite{lovelace2019geocomputation}.

\begin{figure}[!h]
\centering
\begin{python}
def get_utm_crs_from_geodataframe(gdf):
    """
    Determine the appropriate UTM CRS for a given GeoDataFrame.
    
    Parameters:
        gdf: GeoDataFrame with the input geometries
    
    Returns:
        utm_crs: The EPSG code for the appropriate UTM zone
    """
    centroid = gdf.unary_union.centroid
    lon, lat = centroid.x, centroid.y
    # Get the UTM zone
    utm_zone = int((lon + 180) // 6) + 1
 
    # Construct the EPSG code 
    if lat >= 0: #north hemisphere
        epsg_code = 32600 + utm_zone  
    else: #south hemisphere
        epsg_code = 32700 + utm_zone
    
    return epsg_code
\end{python}
\vspace{- 10pt}
\caption{Python Function to calculate the UTM coordinate from the Geopandas \emph{GeoDataFrame}.}
\label{fig:code_coord}
\end{figure}

\newpage
\section{Ground truth}
\label{sup:ground_truth}

\subsection{Landuse information to residential or non-residential}
For each dataset, we list the original land use information and its respective classification as residential (``RES''), non-residential (``NON\_RES''), or mixed-use (``N/A''). Note that for those that are set to a primary use but can allow mixed-use, we set them as ``N/A''.

\begin{itemize}
    \item \textbf{Minneapolis and St. Paul:} 
    \begin{itemize}
        \item Agricultural: N/A;
        \item Airport or Airstrip: NON\_RES;
        \item Extractive: N/A;
        \item Farmstead: N/A;
        \item Golf Course: NON\_RES;
        \item Industrial or Utility: NON\_RES;
        \item Institutional: NON\_RES;
        \item Major Highway: N/A;
        \item Major Railway: N/A;
        \item Manufactured Housing Park: RES;
        \item Mixed Use Commercial: NON\_RES; 
        \item Mixed Use Industrial: NON\_RES; 
        \item Mixed Use Residential: N/A;
        \item Multifamily: RES;
        \item Office: NON\_RES;
        \item Open Water: N/A;
        \item Park, Recreational, or Preserve: N/A;
        \item Retail and Other Commercial: NON\_RES,
        \item Seasonal/Vacation: N/A; 
        \item Single Family Attached: RES;
        \item Single Family Detached: RES;
        \item Undeveloped: N/A.
    \end{itemize}
    The column used to extract the land use tag is ``DESC2020''.
    \item \textbf{Baltimore, MD:} 
    \begin{itemize}
        \item AGRICULTURAL VACANT: N/A;
        \item AGRICULTURE: N/A;
        \item AIRPORT: NON\_RES;
        \item ASSISTED LIVING FACILITY: N/A;
        \item CEMETARY W/O PLACE OF WORSHIP: N/A;
        \item COLLEGE: NON\_RES;
        \item COMMERCIAL: NON\_RES;
        \item COUNTY OPEN SPACE: N/A;
        \item COUNTY PARK: N/A;
        \item COUNTY SENIOR CENTER: N/A;
        \item ELECTRIC, GAS, TELECOMMUNICATIONS UTILITY: N/A;
        \item FIRE FACILITY: NON\_RES;
        \item FURTHER REVIEW: N/A;
        \item HOA/COA/DEVELOPER/MULTIFAMILY MGMT: RES;
        \item HOSPITAL: NON\_RES;
        \item INDUSTRIAL: NON\_RES;
        \item LANDFILL: N/A;
        \item LIBRARY: NON\_RES;
        \item MISC. GOVERNMENT-PUBLIC: NON\_RES;
        \item MISC. INSTITUTION-PRIVATE: NON\_RES;
        \item MIXED OFFICE/INDUSTRIAL: NON\_RES;
        \item MIXED OFFICE/INDUSTRIAL/RETAIL: NON\_RES,
        \item MIXED OFFICE/RETAIL: NON\_RES,
        \item MIXED RESIDENTIAL/OFFICE/RETAIL: N/A;
        \item MULTI SFD: RES;
        \item MULTIFAMILY: RES;
        \item NON-COUNTY PARCEL: RES;
        \item OFFICE: NON\_RES;
        \item OTHER GOVERNMENT OPEN SPACE: N/A;
        \item OTHER PRIVATE OPEN SPACE: N/A;
        \item OTHER PUBLIC PARK: N/A;
        \item PARK AND RIDE: N/A;
        \item PERMANENT EASEMENT: N/A;
        \item PIPELINE: N/A;
        \item PLACE OF WORSHIP: N/A;
        \item POLICE FACILITY: NON\_RES;
        \item PRIVATE SCHOOL: NON\_RES;
        \item PRIVATELY OWNED GOLF COURSE: NON\_RES;
        \item PUBLIC SCHOOL OR SCHOOL SITE: NON\_RES;
        \item PUBLICLY OWNED GOLF COURSE: NON\_RES;
        \item RAIL: N/A;
        \item RESERVOIR PROPERTY: N/A;
        \item ROAD: N/A;
        \item RURAL RESIDENTIAL SFD: N/A;
        \item SFA: RES;
        \item SFD: RES;
        \item SFSD: RES;
        \item STATE PARK: N/A;
        \item STORM DRAINAGE: N/A;
        \item UNBUILDABLE/ENVIRONMENTALLY CONSTRAINED: N/A;
        \item VACANT: N/A;
        \item WATER: N/A;
        \item WATER OR SEWER UTILITY: N/A.
    \end{itemize}
    The column used to extract the land use tag is: ``GIS\_LU\_COD''.
    \item \textbf{Boulder, CO:}
    \begin{itemize}
        \item Agricultural: N/A;
        \item Commercial: NON\_RES;
        \item Foundation/Ruin: N/A;
        \item Garage/Shed: N/A;
        \item Industrial: NON\_RES;
        \item Medical: NON\_RES;
        \item Misc: N/A;
        \item Parking Structure: N/A;
        \item Public: NON\_RES;
        \item Public Safety: NON\_RES;
        \item Religious: NON\_RES;
        \item Residential: RES,
        \item School: NON\_RES;
        \item Tank: N/A.
    \end{itemize}
    The column used to extract the land use tag is ``BLDGTYPE''.
    \item \textbf{Fairfax, VA:}
    \begin{itemize}
        \item High-density Residential: RES;
        \item Low-density Residential: RES;
        \item Medium-density Residential: RES;
        \item Agricultural: N/A;
        \item Commercial: NON\_RES;
        \item Industrial, light and heavy: NON\_RES;
        \item Institutional: NON\_RES;
        \item Open land, not forested or developed: N/A;
        \item Public: NON\_RES;
        \item Recreation: NON\_RES;
        \item Surface water: N/A;
        \item Utilities: N/A;
        \item Industrial: NON\_RES;
        \item Institutional - General: NON\_RES;
        \item Institutional - Government: NON\_RES;
        \item Mixed-Use Residential/Commercial: N/A;
        \item Open Space - Private: N/A;
        \item Open Space - Public: N/A;
        \item Residential - Multifamily: RES;
        \item Residential - Single Attached: RES;
        \item Residential - Single Detached: RES;
        \item Vacant: N/A;
        \item None: N/A (None type).
    \end{itemize}
    The column used to extract the land use tag is ``CATEG'' in the case of the county and ``ELU'' in the case of Fairfax City.
    \item \textbf{Hanover, VA:}
    \begin{itemize}
        \item A-1: N/A;
        \item AR-1: N/A;
        \item AR-2: N/A;
        \item AR-6: N/A;
        \item B-1: NON\_RES;
        \item B-2: NON\_RES;
        \item B-3: NON\_RES;
        \item B-4: NON\_RES;
        \item B-O: NON\_RES;
        \item HE: N/A;
        \item M-1: NON\_RES;
        \item M-2: NON\_RES;
        \item M-3: NON\_RES;
        \item MX: N/A;
        \item O-S: NON\_RES;
        \item PMH: RES;
        \item PSC: NON\_RES;
        \item PUD: N/A;
        \item R-1: RES;
        \item R-2: RES;
        \item R-3: RES;
        \item R-4: RES;
        \item R-5: RES;
        \item R-6: RES;
        \item RC: N/A;
        \item RM: RES;
        \item RO-1: N/A;
        \item RR-1: N/A;
        \item RRC: N/A;
        \item RS: RES;
        \item See Map: N/A;
        \item None: N/A.
    \end{itemize}
    The column used to extract the land use tag is: ``ZONING\_LIS''.
    \item \textbf{Mecklenburg, NC:}
    \begin{itemize}
        \item 100 YEAR FLOOD PLAIN - AC: N/A;
        \item 100 YEAR FLOOD PLAIN - LT: N/A;
        \item AGRICULTURAL - COMMERCIAL PRODUCTION: N/A;
        \item AIR RIGHTS PARCEL: N/A;
        \item AIRPORT: NON\_RES;
        \item AUTO SALES AND SERVICE: NON\_RES;
        \item BANK: NON\_RES;
        \item BILL BOARD: N/A;
        \item BUFFER STRIP: N/A;
        \item CAR WASH: NON\_RES;
        \item CELL TOWER: NON\_RES;
        \item CHURCH: NON\_RES;
        \item CLUB, LODGES, UNION HALL, SWIM CLUB: NON\_RES;
        \item COLLEGE - PUBLIC: NON\_RES;
        \item COMMERCIAL: NON\_RES;
        \item COMMERCIAL COMMON AREA: NON\_RES;
        \item COMMERCIAL CONDOMINIUM: NON\_RES;
        \item COMMERCIAL CONDOMINIUM COMMON AREA: NON\_RES;
        \item COMMERCIAL SERVICE(LAUNDRY,TV,RADIO,ETC): NON\_RES;
        \item COMMERCIAL WATER FRONTAGE: NON\_RES;
        \item CONDO AFFORDABLE HOUSING: RES;
        \item CONDOMINIUM: RES;
        \item CONDOMINIUM COMMON AREA: RES;
        \item CONDOMINIUM HIGH RISE: RES;
        \item CONDOMINIUM WATER FRONTAGE: RES;
        \item CONDOMINIUM WATER VIEW: RES;
        \item CONSERVATION - AGRICULTURAL COMM: N/A;
        \item CONSERVATION - FORESTRY COMM: N/A;
        \item CONSERVATION - WILDLIFE: N/A;
        \item CONSERVATION - WOODLAND EXCESS AC: N/A;
        \item CONVENIENCE STORE: NON\_RES;
        \item CONVIENCE/FAST FOOD STORE: NON\_RES;
        \item COUNTRY CLUB: NON\_RES;
        \item DAY CARE CENTER: NON\_RES;
        \item DEPARTMENT STORE: NON\_RES;
        \item ENVIRONMENTAL HAZARD: NON\_RES;
        \item FAST FOOD: NON\_RES;
        \item FIRE DEPARTMENT: NON\_RES;
        \item FLUM/SWIM FLOODWAY (NO BUILD ZONE): N/A;
        \item FOREST - COMMERCIAL PRODUCTION: N/A;
        \item FUNERAL (MORTUARY, CEMETERY, CREMATORIUM, MAUS): NON\_RES;
        \item GOLF COURSE CLASS 1 - CHAMPIONSHIP: NON\_RES;
        \item GOLF COURSE CLASS 2 - PRIVATE CLUB: NON\_RES;
        \item GOLF COURSE CLASS 3 - SEMI-PRIVATE \& MUNICIPAL: NON\_RES;
        \item GOLF COURSE CLASS 4 - MINIMUM QUALITY: NON\_RES;
        \item GREENWAY TRAIL: N/A;
        \item HABITAT FOR HUMANITY: N/A;
        \item HOME FOR THE AGED: RES;
        \item HORTICULTURAL - COMMERCIAL PRODUCTION: N/A;
        \item HOSPITAL, PRIVATE: NON\_RES;
        \item HOSPITALS - PUBLIC: NON\_RES;
        \item HOTEL/MOTEL $<$ 7 FLOORS: NON\_RES;
        \item HOTEL/MOTEL $>$ 6 FLOORS: NON\_RES;
        \item INDUSTRIAL: NON\_RES;
        \item INDUSTRIAL COMMON AREA: NON\_RES;
        \item INDUSTRIAL PARK: NON\_RES;
        \item INSTITUTIONAL: NON\_RES;
        \item ISLAND: N/A;
        \item LABORATORY / RESEARCH: NON\_RES;
        \item LEASEHOLD INTEREST: NON\_RES;
        \item LIGHT MANUFACTURING: NON\_RES;
        \item LUMBER YARD: NON\_RES;
        \item MARINA LAND: NON\_RES;
        \item MEDICAL CONDOMINIUM: NON\_RES;
        \item MEDICAL CONDOMINIUM COMMON AREA: NON\_RES;
        \item MEDICAL OFFICE: NON\_RES;
        \item MINI WAREHOUSE: NON\_RES;
        \item MINIATURE GOLF COURSES/DRIVING RANGE: NON\_RES;
        \item MINING: N/A;
        \item MOBILE HOME HS: RES;
        \item MOBILE HOME PARK: N/A;
        \item MOBILE HOME SUBDIVISION: N/A;
        \item MULTI FAMILTY AFFORDABLE HOUSING: RES;
        \item MULTI FAMILY: RES;
        \item MULTI FAMILY COMMON AREA: RES;
        \item MULTI FAMILY DUPLEX/TRIPLEX: RES;
        \item MULTI FAMILY GARDEN: N/A;
        \item MULTI FAMILY HIGH RISE: RES;
        \item MULTI FAMILY TOWNHOUSE: RES;
        \item MULTI FAMILY WATER ACCESS: N/A;
        \item MUNICIPAL AIRPORT: NON\_RES;
        \item MUNICIPAL EDUCATION: NON\_RES;
        \item NEW PARCEL: N/A;
        \item NO LAND INTEREST: N/A;
        \item NURSING HOME: RES;
        \item OFFICE: NON\_RES;
        \item OFFICE CONDOMINIUM: NON\_RES;
        \item OFFICE CONDOMINIUM COMMON AREA: NON\_RES;
        \item OFFICE HIGH RISE - $>$ 6 STORIES: NON\_RES;
        \item OTHER COUNTY PROPERTY: NON\_RES;
        \item OTHER FEDERAL: NON\_RES;
        \item OTHER MUNICIPAL: NON\_RES;
        \item PACKING PLANT: NON\_RES;
        \item PARKING: N/A;
        \item PATIO HOME: N/A;
        \item PATIO HOME - WATERFRONT: N/A;
        \item PETROLEUM AND GAS: N/A;
        \item PVT Owned RR with Rail ROW: N/A;
        \item R101: N/A;
        \item REC AREA: N/A;
        \item RESERVED PARCEL: N/A;
        \item RESIDENTIAL AFFORDABLE HOUSING: RES;
        \item RESTAURANT: NON\_RES;
        \item RIGHT OF WAY: N/A;
        \item ROADWAY CORRIDOR: N/A;
        \item RURAL HOMESITE: N/A;
        \item SCHOOL - PUBLIC: NON\_RES;
        \item SCHOOL,COLLEGE, PRIVATE: NON\_RES;
        \item SERVICE GARAGE: NON\_RES;
        \item SERVICE STATION: NON\_RES;
        \item SHOPPING CENTER - MALL: NON\_RES;
        \item SHOPPING CENTER - STRIP: NON\_RES;
        \item SINGLE FAMILY RESIDENTIAL: RES;
        \item SINGLE FAMILY RESIDENTIAL - ACREAGE: RES;
        \item SINGLE FAMILY RESIDENTIAL - COMMON: RES;
        \item SINGLE FAMILY RESIDENTIAL - GOLF: RES;
        \item SINGLE FAMILY RESIDENTIAL - RIVER: RES;
        \item SINGLE FAMILY RESIDENTIAL - WATER VIEW: RES;
        \item SINGLE FAMILY RESIDENTIAL - WATERFRONT: RES;
        \item SINGLE FAMILY RESIDENTIAL GATED COMMUNITY: RES;
        \item SINGLE FAMILY RESIDENTIAL MINI FARM/ESTATE: RES;
        \item STATE PROP: N/A;
        \item SUBMERGED LAND, RIVERS AND LAKES: N/A;
        \item SUPERMARKET: NON\_RES;
        \item TOWN HOUSE  GOLF COURSE FRONTAGE: RES;
        \item TOWN HOUSE  SFR: RES;
        \item TOWN HOUSE  WATER ACCESS: RES;
        \item TOWN HOUSE  WATER FRONTAGE: RES;
        \item TOWN HOUSE COMMON AREA: RES;
        \item TOWNHOUSE AFFORDABLE HOUSING: RES;
        \item TRUCK TERMINAL: NON\_RES;
        \item UNSUITABLE FOR SEPTIC: N/A;
        \item USE VALUE HOMESITE: N/A;
        \item UTILITY (GAS, ELECTRIC, TELEPHONE, TELEGRAPH, RAIL: N/A;
        \item UTILITY EASEMENT: N/A;
        \item UTILITY/P: N/A;
        \item WAREHOUSE CONDOMINIUM: NON\_RES;
        \item WAREHOUSE CONDOMINIUM COMMON AREA: NON\_RES;
        \item WAREHOUSING: NON\_RES;
        \item WASTELAND, SLIVERS, GULLIES, ROCK OUTCROP: N/A,
        \item WATER PLANT: N/A;
        \item WATER RETENTION POND: N/A;
        \item WELL LOT: N/A;
        \item WETLAND: N/A;
        \item WOODLAND - EXCESS ON AG PCL: N/A.
    \end{itemize}
    The column used to extract the landuse tag is: ``landuse\_de''.
\end{itemize}

\section{No sheds and garages}
\label{sup:no_shads}
Here are the results of the comparison to the ground truth, where we exclude the sheds and garages. Table~\ref{tab:msa_results_no_shed} refers to Minneapolis and St. Paul, and Table~\ref{tab:us_regions_no_shed} refers to the other regions of the U.S.

\begin{table*}[!ht]
\centering
\begin{tabular}{|c|c|c|c|c|c|}
\hline
\textbf{County}             &\textbf{Class} &\textbf{Precision}&\textbf{Recall}&\textbf{F1-Score}&\textbf{Avg. F1-Score} \\ \hline \hline
\multirow{2}{*}{Anoka, MN}  &non-residential&   0.99    &  0.72  &   0.83   &  \multirow{2}{*}{0.89} \\ \cline{2-5} 
                            &  residential  &   0.90    &  1.00  &   0.95   &  \\ \hline
\multirow{2}{*}{Carver, MN} &non-residential&   0.96    &  0.75  &   0.84   &  \multirow{2}{*}{0.91} \\ \cline{2-5}
                            &  residential  &   0.96    &  1.00  &   0.98   &     \\ \hline
\multirow{2}{*}{Dakota, MN} &non-residential&   0.98    &  0.77  &   0.86   &  \multirow{2}{*}{0.93} \\ \cline{2-5} 
                            &  residential  &   0.98    &  1.00  &   0.99   &     \\ \hline
\multirow{2}{*}{Hennepin, MN} &non-residential& 0.97    &  0.77  &   0.86   &  \multirow{2}{*}{0.92} \\ \cline{2-5} 
                            &  residential  &   0.97    &  1.00  &   0.98   &     \\ \hline
\multirow{2}{*}{Ramsey, MN}  &non-residential&   0.97    &  0.63  &   0.76   &  \multirow{2}{*}{0.86} \\ \cline{2-5} 
                            &  residential  &   0.94    &  1.00  &   0.97   &     \\ \hline
\multirow{2}{*}{Scott, MN}   &non-residential&   0.98    &  0.70  &   0.81   &  \multirow{2}{*}{0.89} \\ \cline{2-5} 
                            &  residential  &   0.95    &  1.00  &   0.97   &     \\ \hline
\multirow{2}{*}{Washington, MN} &non-residential&0.98   &  0.76  &   0.86   &  \multirow{2}{*}{0.92} \\ \cline{2-5} 
                            &  residential  &   0.96    &  1.00  &   0.98   &     \\ \hline
\end{tabular}
\caption{Prediction results for the Minneapolis and St. Paul area excluding sheds and garages.}
\label{tab:msa_results_no_shed}
\end{table*}

\begin{table*}[!ht]
\centering
\begin{tabular}{|c|c|c|c|c|c|}
\hline
\textbf{Region}                     &\textbf{Class} &\textbf{Precision}&\textbf{Recall}&\textbf{F1-Score}&\textbf{Avg. F1-Score} \\ \hline \hline
\multirow{2}{*}{Baltimore, MD}  &non-residential&   0.94    &  0.83  &   0.88   &  \multirow{2}{*}{0.94} \\ \cline{2-5} 
                                    &  residential  &   0.99    &  1.00  &   0.99   &  \\ \hline 
\multirow{2}{*}{Boulder, CO$^{*}$} &non-residential&   0.85    &  0.70  &   0.77   &  \multirow{2}{*}{0.88} \\ \cline{2-5} 
                                    &  residential  &   0.97    &  0.99  &   0.98   &  \\ \hline 
\multirow{2}{*}{Fairfax, VA$^{**}$}    &non-residential&   0.96    &  0.83  &   0.89   &  \multirow{2}{*}{0.94} \\ \cline{2-5} 
                                    &  residential  &   0.99    &  1.00  &   0.99   &  \\ \hline
                                    
\multirow{2}{*}{Hanover, VA}    &non-residential&   0.97    &  0.62  &   0.75   &  \multirow{2}{*}{0.87} \\ \cline{2-5} 
                                    &  residential  &   0.97    &  1.00  &   0.98   &  \\ \hline 
                                    
\multirow{2}{*}{Mecklenburg, NC}&non-residential&  0.92  &  0.74  &   0.82   &  \multirow{2}{*}{0.91} \\ \cline{2-5} 
                                    &  residential  &   0.98    &  1.00  &   0.99   &  \\ \hline 

\end{tabular}
\caption{Prediction results for other regions of the U.S. excluding sheds and garages. Here, Boulder$^*$ is a city, Fairfax$^{**}$ is both the county and the town of Fairfax, and the others are counties.}
\label{tab:us_regions_no_shed}
\end{table*}

\section{Proportion of buildings without annotations}
\label{sup:map}
\begin{figure*}[!ht]
    \centering
    \includegraphics[width=0.65\textwidth]{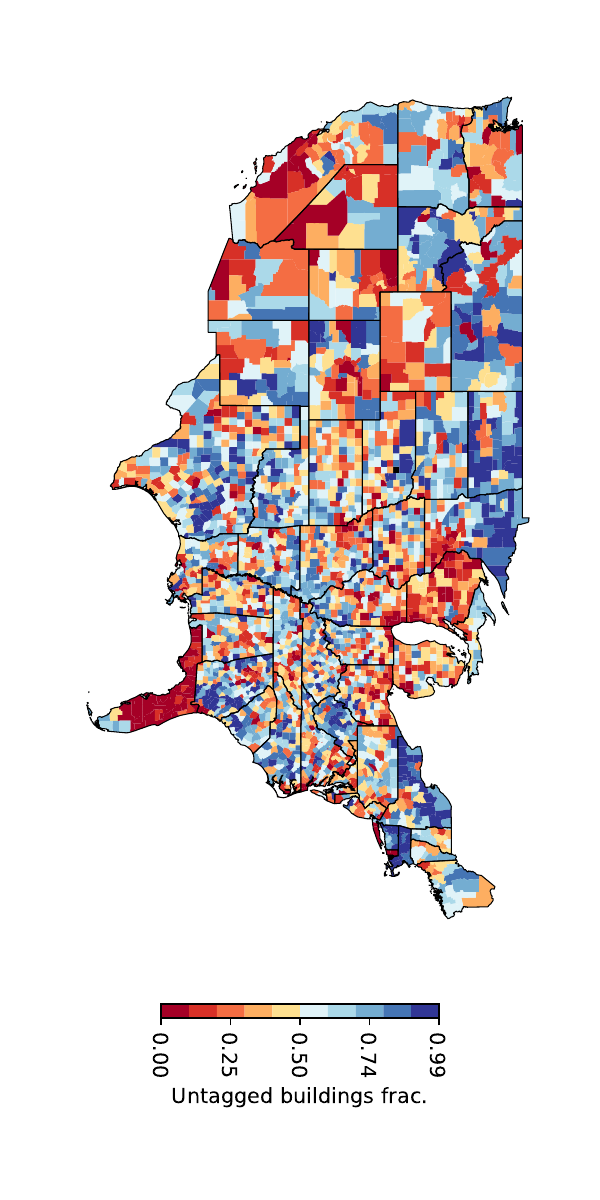}
    
    \caption{Map of the contiguous U.S. with the proportion of untagged buildings per county or equivalent region. The color black indicates a county for which data are not available in OSM (Wheeler County, NE).}
    \label{fig:map}
\end{figure*}

Figure~\ref{fig:map} shows the proportion of untagged buildings per county in the contiguous U.S.

\section{Select particular footprints}
\label{sec:shed_garage}
Figure~\ref{fig:code_shed_garage} shows the Python code to select only the footprints that are not sheds, garages, and parking lots.

\begin{figure}[!h]
\centering
\begin{python}
    ...
    delete = ["building:shed", "building:garage", "building:garages", "building:parking"]
    buildings_df = buildings_df[~buildings_df['tag used'].isin(delete)]
    ...
\end{python}
\vspace{- 10pt}
\caption{\textbf{Code to select only the footprints that are not sheds, garages, and parking lots.} Python code to remove the rows of sheds, garages, and parking lots from the Geopandas \emph{GeoDataFrame}.}
\label{fig:code_shed_garage}
\end{figure}

\end{document}